# Irreversible Diagonalization of Mechanical Quantities and the EPR Paradox

Tao Liu

*Coherent Electron Quantum Optics Research Center State Key Laboratory of Environment-friendly Energy Materials, Southwest University of Science and Technology, Mianyang, Sichuan 621010, China, and*

*School of Science, Southwest University of Science and Technology, Mianyang, Sichuan 621010, China*

The closure relation of quantum mechanical projection operators is not entirely true; it can be strictly falsified under unitary transformations in Fock states. The angular momentum $J_x$, $J_y$ and $J_z$ are simultaneously diagonalized under the orthonormal set $\{|\varphi_n\rangle\}$ of continuous rotation transformations in Fock states. $\{|\varphi_n\rangle\}$'s time reversal $\{\mathbf{T}|\varphi_n\rangle\}$ is the zero point of coordinates **q** and momentum **p**, and its arbitrary translation transformation $\{\mathfrak{D}\mathbf{T}|\varphi_n\rangle\}$ diagonalizes both coordinates and momentum simultaneously. The abstract representation of the Dirac state vector implies the symmetry breaking of the non-Abelian group unit matrix $\{\mathbf{U^H U} \neq \mathbf{UU^H}\}$. The EPR paradox is merely a fallacy under the reversible diagonalization of physical reality; it is resolved under irreversible diagonalization.

## I. INTRODUCTION

In 1927, Werner Heisenberg heuristically described the uncertainty principle: the act of measurement inevitably disturbs the motion state of the measured particle, thus generating uncertainty $\Delta x \Delta p \approx h$ [1]. That same year, Earl Kennard first proved the inequality $\Delta x \Delta p \geq \hbar/2$ [2]. By 1929, Howard Robertson gave the general form of the uncertainty relation: the fluctuations $\Delta A$ and $\Delta B$ of any two mechanical quantities A and B in any state must satisfy:

$$\Delta A \Delta B \geq \frac{1}{2}\left|\overline{[A, B]}\right| \qquad (1.1)$$

where the right side of the inequality is the absolute value of the average of the commutation relation of A and B [3].

According to Robertson's proven uncertainty relation, if two mechanical quantities A and B do not commute, then generally $\Delta A$ and $\Delta B$ cannot both be zero (i.e., A and B cannot both have definite measured values). Or generally speaking, they cannot have a common eigenstate.

In 1935, Einstein, Podolsky and Rosen pointed out that the quantum mechanical description of physical reality provided by the wave function is incomplete [4]. They proposed a criterion for completeness: "Every element of the physical reality must have a counterpart in the physical theory." They also provided a criterion for physical reality: "If, without disturbing a system in any way, one can predict with certainty (probability equal to 1) the value of a physical quantity, then there exists an element of physical reality corresponding to this physical quantity."

According to these conventions, EPR gave an example: if the state $|\psi\rangle$ of a system is an eigenfunction of an observable A with eigenvalue $\alpha$, then there exists an element of physical reality corresponding to the physical quantity A. But if $|\psi\rangle$ is not an eigenfunction, one cannot say that A has a specific value. Since the position and momentum operators do not commute in quantum

mechanics, when the momentum of a particle is known, its position is not a physical reality, and vice versa. From this, they deduced [4]:

1. Either the quantum mechanical description of physical reality provided by the wave function is incomplete;

2. Or, when the operators corresponding to two physical quantities do not commute, these two quantities cannot both be elements of physical reality. This is the famous EPR paradox.

Bohr attempted to refute the EPR argument by pointing out that the epistemological criterion of physical reality is ambiguous. He replaced the concept of unavoidable physical disturbance caused by measurement with the notion that the measured microscopic object and the measuring instrument form an indivisible whole. Bohr argued that the criterion for elements of reality should consider both the measuring instrument and the measured particle [5].

Einstein insisted that the epistemological criterion of physical reality is independent of experiments, while Bohr maintained that the criterion should consider both the measuring instrument and the measured particle. Neither convinced the other throughout their lives.

Logically, the cause of the EPR paradox is the impossibility of two non-commuting mechanical quantities having a common eigenstate under an orthogonal normalization complete set. This seems to be a theorem that can be strictly proven [Appendix 1]. As Newton said, "In experimental physics, theorems derived by induction from phenomena should be considered as true or nearly true until other phenomena arise that might correct them or provide exceptions" [6].

Unfortunately, the basic premise of "two non-commuting mechanical quantities cannot have a common eigenstate" is that the projection operator in quantum mechanics is an identity operator [7]. This cannot be completely verified but can be strictly falsified. The reason lies in the long-standing tendency to regard Dirac state vectors as abstract representations of wave functions, ignoring the important physics implied by the non-Abelian group under the condition $\{\mathbf{U^H U} \neq \mathbf{U U^H}\}$.

This paper first elaborates from the perspective of matrix theory in linear algebra, explaining the symmetry breaking caused by extending unitary matrices in complex space to those expressed by Dirac state vectors. It points out that the **U** matrix composed of orthogonal normalized bases expressed by Dirac kets and its transpose conjugate $\mathbf{U^H}$ is a gauge $\mathbf{U^H U = I}$. When $\mathbf{U U^H = U^H U}$, the reversible diagonalization $\mathbf{U^H F U}=\mathrm{diag}(f_1,\cdots,f_n)$ under the Abelian group gives the first kind of constant of motion. When $\mathbf{U U^H \neq U^H U}$, the irreversible diagonalization $\mathbf{U^H F U}=\mathrm{diag}(f_1',\cdots,f_n')$ under the non-Abelian group gives the second kind of constant of motion. We then state that the proof of the projection operator as an identity operator $\mathbf{U U^H = I}$ is always circular and can be strictly falsified under the unitary transformation of Fock states. As examples, we provide the common eigenstates of angular momenta $J_x$, $J_y$ and $J_z$ and their squares $J_x^2$, $J_y^2$ and $J_z^2$, as well as the common eigenstates of coordinate **q** and momentum **p** and their squares $\mathbf{q}^2$ and $\mathbf{p}^2$. Their eigenvalues still satisfy Robertson's uncertainty relation. The second kind of constant of motion is closely related to quantum phase transitions. Finally, we point out that the necessary condition for completeness and the sufficient condition for elements of physical reality proposed by Einstein, Podolsky, and Rosen are valid under the orthogonal normalized complete set expressed by Dirac state vectors. The quantum mechanical description of physical reality provided by the wave function remains complete.

## II. EXTENSION OF UNITARY MATRICES IN COMPLEX SPACE

According to the matrix theory of linear algebra, the definition and basic properties of unitary

matrices and the basic theorem of Hermitian matrices are as follows:

Definition of Unitary Matrix: Let $\mathbf{A} \in \mathbf{C}^{n \times n}$, if $\mathbf{AA^H=I}$ (where $\mathbf{A^H}$ is the conjugate transpose of $\mathbf{A}$), then A is called a unitary matrix.

Basic Properties of Unitary Matrices: 1. $\mathbf{A^H = A^{-1}}$. 2. $\mathbf{AA^{-1} = A^{-1}A = I}$. 3. $|\mathbf{detA}| = 1$.

Basic Theorem of Hermitian Matrices:

Let $\mathbf{F}$ be an n-order Hermitian matrix, then there exists a unitary matrix $\mathbf{A}$ such that $\mathbf{A^{-1}FA}$ is a diagonal matrix $diag(f_1, \cdots, f_n)$, where $f_1, \cdots, f_n$ are the n eigenvalues of $\mathbf{F}$, and the n columns of the unitary matrix are composed of n linearly independent eigenvectors.

Dirac state vector representation of unitary matrix The orthonormal state vector $|\varphi_i\rangle$ and headjoint ranspose $\langle\varphi_i|$ must form two diagonal matrices $\mathbf{U}$ and $\mathbf{U^H}$ that is, the assumption $\langle\varphi_i|\varphi_j\rangle = \delta_{ij}$, must hold

$$\mathbf{U} = diag(|\varphi_1\rangle, \cdots, |\varphi_n\rangle)$$
$$\mathbf{U^H} = diag(\langle\varphi_1|, \cdots, \langle\varphi_n|) \quad (2.1)$$

$\mathbf{U^H}$ formally satisfies the definition of a unitary matrix $\mathbf{U^H(U^H)^H = I}$. It is the extension of the unitary matrix in the scalar space (matrix elements $\in$ C) to the Dirac state vector space (matrix elements $\in |\ \rangle$). The extended unitary matrix exhibits symmetry breaking compared to the unitary matrix in scalar space: $\mathbf{U^HU \neq UU^H}$ (Compare $\mathbf{A^HA=AA^H}$). Including Dirac state vectors $\mathbf{U^H}$ and $\mathbf{U}$ which makes the square matrix $\mathbf{U^HU}$ and $\mathbf{UU^H}$ not the same. Because

$$\mathbf{U^HU} = diag(\langle\varphi_1|\varphi_1\rangle, \cdots, \langle\varphi_n|\varphi_n\rangle) = \mathbf{I}$$
$$\mathbf{UU^H} = diag(|\varphi_1\rangle\langle\varphi_1|, \cdots, |\varphi_n\rangle\langle\varphi_n|) \quad (2.2)$$

the square matrix $\mathbf{U^HU}$ has matrix elements that are structureless scalars 1, while the square matrix $\mathbf{UU^H}$ has matrix elements that are structured submatrices $|\varphi_i\rangle\langle\varphi_i|$. In other words, the properties of unitary matrices in the complex space of linear algebra cannot be simply generalized to the Dirac state vector space. Expressed by the Dirac state vector $\mathbf{U^HU}$ is a gauge, while $\mathbf{UU^H}$ is the projection operator.

2. Two types of transformations of mechanical quantities

2.1 Reversible transformations under the Abelian group ($\mathbf{UU^H=U^HU}$)

If $\mathbf{UU^H = U^HU}$ for the mechanical quantity $\mathbf{F}$, if $\langle\varphi_m|\mathbf{F}|\varphi_n\rangle = \delta_{mn}f_n$, that is, $\mathbf{U^HFU} = diag(f_1, \cdots, f_n)$, then the diagonalization of $\mathbf{F}$ is reversible:

$$\mathbf{F} = \mathbf{UU^HFUU^H} = \mathbf{U}diag(f_1, \cdots, f_n)\mathbf{U^H} \quad (2.3)$$

The reversible diagonalization under the Abelian group gives the first type of motion constant $\{f_i\}$ for the mechanical quantity $\mathbf{F}$. At this time, the eigenvalue of the mechanical quantity $\{f_i\}$ has no fluctuation:

$$\overline{(\Delta \mathbf{F})^2} = \mathbf{U^HF^2U} - \mathbf{U^HFUU^HFU} = \mathbf{0} \quad (2.4)$$

The mechanical quantity F is equivalent to its reversible transformation $\mathbf{U^HFU}$: $\mathbf{F} = \mathbf{UFU^H}$.

2.2 Irreversible transformation under the non-Abelian group ($\mathbf{UU^H \neq U^HU}$)

If $\mathbf{UU^H \neq U^HU}$, for the mechanical quantity F, if $\langle\varphi_m|\mathbf{F}|\varphi_n\rangle = \delta_{mn}f'_n$, that is, $\mathbf{U^HFU} = diag(f'_1, \cdots, f'_n)$, Therefore, the diagonalization of $\mathbf{F}$ is irreversible:

$$\mathbf{F} \neq \mathbf{UU^HFUU^H} = \mathbf{U}diag(f'_1, \cdots, f'_n)\mathbf{U^H} \quad (2.5)$$

The irreversible diagonalization under the non-Abelian group gives the second type of motion constant of the mechanical quantity $\{f'_i\}$. At this time, the eigenvalues of the mechanical quantity $\{f'_i\}$ have fluctuations:

$$\overline{(\Delta \mathbf{F})^2} = \mathbf{U^HF^2U} - \mathbf{U^HFUU^HFU} \neq \mathbf{0} \quad (2.6)$$

The mechanical quantity $\mathbf{F}$ is similar to its irreversible transformation $\mathbf{U^HFU}$: $\mathbf{F} \backsim \mathbf{U^HFU}$.

# III. CLOSURE RELATION OF PROJECTION OPERATORS $\sum_i |u_i\rangle\langle u_i| = I$ REFUTATION

So far, people have regarded the sum of the orthonormal basis as $\mathbf{U}\mathbf{U}^H = \sum_i |\varphi_i\rangle\langle\varphi_i|$ as the identity operator $\mathbf{U}\mathbf{U}^H = \mathbf{I}$, that is for any state $|\psi\rangle$, it holds that $\mathbf{U}\mathbf{U}^H|\psi\rangle = \mathbf{I}|\psi\rangle = |\psi\rangle$ Unfortunately, a careful examination of the proof of the projection operator $\mathbf{U}\mathbf{U}^H = \mathbf{I}$ under arbitrary states reveals that: people have assumed a priori that $\mathbf{U}\mathbf{U}^H = \mathbf{I}$ to be true before proving this proposition. In fact, this is not true for any state $|\psi\rangle$! $\mathbf{U}\mathbf{U}^H = \mathbf{I}$ can be strictly falsified. To clarify this point more clearly, we first present Cohen-Tannoudji's proof regarding the closure relation $\sum_i |\varphi_i\rangle\langle\varphi_i| = \mathbf{I}$[7] and point out its a priori assumptions, and then take the Fock state as an example to discuss its closure relation. $\sum_n |n\rangle\langle n| = \mathbf{I}$ Falsifiability.

3.1 Cohen·Tannoudji on the closure relation $\sum_i |u_i\rangle\langle u_i| = \mathbf{I}$'s circular argument[7]

"a. Orthogonal normalization relation
We say that the discrete set of right vectors $\{|u_i\rangle\}$ is orthonormal, provided that the right vectors in the set satisfy the following orthonormal relation

$$\langle u_i|u_j\rangle = \delta_{ij} \tag{C-1}$$

b. Closure relation The discrete set $\{|u_i\rangle\}$ forms a basis if: ε every right vector in the space $|\psi\rangle$ can be uniquely expressed as $|u_i\rangle$ expanded as:

$$|\psi\rangle = \sum_i c_i |u_i\rangle \tag{C-3}$$

Furthermore, assuming the basis is orthonormal, by multiplying both sides of $\langle u_j|$ by Eq.(C-3) and using Eq.(C-1), we obtain the expression for the component $c_j$:

$$\langle u_j|\psi\rangle = c_j \tag{C-5}$$

Replace Eq.(C-3) in $c_i$ with $\langle u_i|\psi\rangle$, Then we have

$$|\psi\rangle = \sum_i c_i |u_i\rangle = \sum_i \langle u_i|\psi\rangle|u_i\rangle = \sum_i |u_i\rangle\langle u_i|\psi\rangle = \left(\sum_i |u_i\rangle\langle u_i|\right)|\psi\rangle \tag{C-7}$$

[This is because in Eq.(C-7) we can place the number $\langle u_i|\psi\rangle$ to the right vecto $|u_i\rangle$ on the right side].

Thus we see the operator appears: $\sum_i |u_i\rangle\langle u_i|$, applying it to ε any right vector $|\psi\rangle$, we still obtain that right vector $|\psi\rangle$. Since $|\psi\rangle$ is arbitrary, we can obtain:

$$\mathbf{P}_{\{u_i\}} = \sum_i |u_i\rangle\langle u_i| = \mathbf{I} \tag{C-9}$$

In the expression, I represents the ε identity operator in space, and the relation Eq.(C-9) is called the closure relation."

The above is Cohen·Tannoudji's proof regarding discrete closure relations. There is no doubt that the above proof process is impeccable! The problem lies in the premise Eq.(C-3) equation. Indeed, "ε every right vector in the space $|\psi\rangle$ can be uniquely expressed as $|u_i\rangle$ expands". But this is only merely "*possibility*". When we rephrase the question: how can we uniquely expand any state vector $|\psi\rangle$ as $|u_i\rangle$? Then the problem becomes evident: transforming "*possibility*" into "*reality*" necessarily corresponds to a definite operation $\mathbf{P}_{\{u_i\}}|\psi\rangle$. "*possibility*" The strict formulation that should be realized is

$$|\psi\rangle \rightarrow \sum_i c_i |u_i\rangle \implies \mathbf{P}_{\{u_i\}}|\psi\rangle = \sum_i c_i |u_i\rangle \tag{4.1}$$

And expressing "*possibility*" as Eq.(C-3) in the form of $|\psi\rangle = \sum_i c_i |u_i\rangle$ At this point, it has already been implicitly assumed that the projection Operator is the identity operator $\mathbf{P}_{\{u_i\}} = \mathbf{I}$. Thus, the general argument regarding the projection operator $\sum_i |u_i\rangle\langle u_i| = \mathbf{I}$ ultimately falls into an invalid circular argument logically. In fact, the identity operator $\sum_i |u_i\rangle\langle u_i| = I$ can be strictly refuted.

3.2 The falsification of the closure relation $\sum_i |u_i\rangle\langle u_i| = I$
3.2.1 The falsification of the closure relation under the transformation of $\Theta$

Refutation of closure relation under transformation Intro duce the unitary operator $\Theta = e^{\xi b^\dagger c - \xi^* bc^\dagger}$, $\Theta^{-1} = e^{-\xi b^\dagger c + \xi^* bc^\dagger}$, $\Theta\Theta^{-1} = \Theta^{-1}\Theta = 1$. Where $\xi = re^{-i\theta}$. The annihilation and creation operators satisfy $[b, b^\dagger] = 1$, $[c, c^\dagger] = 1$, $[b, c^\dagger] = [c, b^\dagger] = [b^\dagger, c^\dagger] = [b, c] = 0$.

Let Fock state $|n\rangle = \frac{(c^\dagger)^n}{\sqrt{n!}}|0\rangle$ The unitary transformation of is

$$|\varphi_n\rangle = \Theta|n\rangle = \sum_{k=0}^{n} \frac{(C_n^k)^{1/2} \tau^{n-k}}{(1+\tau^*\tau)^{n/2}} |n-k, k\rangle$$
$$\langle\varphi_n| = \langle n|\Theta^{-1} = \sum_{k=0}^{n} \langle k, n-k| \frac{(C_n^k)^{1/2} \tau^{*n-k}}{(1+\tau^*\tau)^{n/2}}$$
(3.1)

Here $\tau = \tan(r) e^{-i\theta}$, $\tau^* = \tan(r) e^{i\theta}$; $C_n^k = \frac{n!}{(n-k)!k!}$. And $\{|\varphi_n\rangle\}$ Satisfies $\langle\varphi_m|\varphi_n\rangle = \delta_{mn}$.

Assumption

$$\sum_{n=0}^{\infty} |n\rangle\langle n| = I \tag{3.2}$$

Must have

$$\Theta\left(\sum_{n=0}^{\infty} |n\rangle\langle n|\right)\Theta^{-1} = \Theta I \Theta^{-1} = I \tag{3.3}$$

That is to say

$$\sum_{n=0}^{\infty} |\varphi_n\rangle\langle\varphi_n| = I \tag{3.4}$$

is the identity operator. Now take

$$|\psi\rangle = (b^\dagger c + bc^\dagger)|\varphi_n\rangle \tag{3.5}$$

must have

$$\langle\psi|\left(\sum_{n=0}^{\infty} |\varphi_n\rangle\langle\varphi_n|\right)|\psi\rangle = \langle\psi|\psi\rangle \tag{3.6}$$

However, it can be strictly concluded that:
Eq.(3.6) Left side of the equation

$$\langle\psi|\left(\sum_{n=0}^{\infty} |\varphi_n\rangle\langle\varphi_n|\right)|\psi\rangle = n^2 \left(\frac{\tau^* + \tau}{1 + \tau^*\tau}\right)^2 \tag{3.6L}$$

Eq.(3.6) Right side of the equation

$$\langle\psi|\psi\rangle = n + n(n-1)\left(\frac{\tau^* + \tau}{1 + \tau^*\tau}\right)^2 \tag{3.6R}$$

Since Eq. (3.6L) $\neq$ Eq. (3.6R), the assumption is invalid! Fock state closure relation $\sum_n |n\rangle\langle n| =$

**I** is falsified under **Θ** transformation.

3.2.2 The falsification of the closure relation under the transformation of **D**

Introduce the unitary operator $\mathbf{D} = e^{-\alpha c^\dagger + \alpha^* c}$, $\mathbf{D}^{-1} = e^{\alpha c^\dagger - \alpha^* c}$, $\mathbf{DD}^{-1} = \mathbf{D}^{-1}\mathbf{D} = 1$;

And the parity operator $\mathbf{\Pi} = e^{i\pi c^\dagger c}$, $\mathbf{\Pi}^{-1} = e^{-i\pi c^\dagger c}$, $\mathbf{\Pi\Pi}^{-1} = \mathbf{\Pi}^{-1}\mathbf{\Pi} = 1$.

The translation transformation of the Fock state $|n\rangle$ is obviously

$$|n\rangle_\alpha = \mathbf{D}|n\rangle = \frac{(c^\dagger + \alpha)^n}{\sqrt{n!}} e^{-\alpha c^\dagger + \alpha^* c}|0\rangle$$

$$_\alpha\langle n| = \langle n|\mathbf{D}^{-1} = \langle 0|e^{\alpha c^\dagger - \alpha^* c}\frac{(c + \alpha^*)^n}{\sqrt{n!}}$$
(3.7)

$\{|n\rangle_\alpha\}$ Satisfies orthogonal normalization $_\alpha\langle m|n\rangle_\alpha = \langle m|\mathbf{D}^{-1}\mathbf{D}|n\rangle = \delta_{mn}$

Assumption

$$\sum_{n=0}^{\infty} |n\rangle\langle n| = \mathbf{I} \quad (3.8)$$

must have

$$\mathbf{D}\left(\sum_{n=0}^{\infty} |n\rangle\langle n|\right)\mathbf{D}^{-1} = \mathbf{DID}^{-1} = \mathbf{I} \quad (3.9)$$

That is to say

$$\sum_{n=0}^{\infty} |n\rangle_{\alpha\alpha}\langle n| = \mathbf{I} \quad (3.10)$$

is the identity operator. Applying this identity operator to the parity operator $\mathbf{\Pi}$ must yield

$$\sum_{j=0}^{\infty} |j\rangle_{\alpha\alpha}\langle j|\mathbf{\Pi}^2 \sum_{k=0}^{\infty} |k\rangle_{\alpha\alpha}\langle k| = \sum_{j=0}^{\infty} |j\rangle_{\alpha\alpha}\langle j|\mathbf{\Pi} \sum_{n=0}^{\infty} |n\rangle_{\alpha\alpha}\langle n|\mathbf{\Pi} \sum_{k=0}^{\infty} |k\rangle_{\alpha\alpha}\langle k| \quad (3.11)$$

Since $\mathbf{\Pi}^2 |k\rangle_\alpha \equiv |k\rangle_\alpha$, therefore

Eq.(3.11) Left side of the equation

$$\sum_{j=0}^{\infty} |j\rangle_{\alpha\alpha}\langle j|\mathbf{\Pi}^2 \sum_{k=0}^{\infty} |k\rangle_{\alpha\alpha}\langle k| = \sum_{j=0}^{\infty} |j\rangle_{\alpha\alpha}\langle j| \sum_{k=0}^{\infty} |k\rangle_{\alpha\alpha}\langle k| = \mathbf{I} \quad (3.11L)$$

Eq.(3.11) Right side of the equation

$$\sum_{j=0}^{\infty} |j\rangle_{\alpha\alpha}\langle j|\mathbf{\Pi} \sum_{n=0}^{\infty} |n\rangle_{\alpha\alpha}\langle n|\mathbf{\Pi} \sum_{k=0}^{\infty} |k\rangle_{\alpha\alpha}\langle k| = \sum_{j,k=0}^{\infty} \sum_{n=0}^{\infty} D_{jn} D_{nk} |j\rangle_{\alpha\alpha}\langle k| \quad (3.11R)$$

Where

$$D_{jn} = {}_\alpha\langle j|e^{i\pi c^\dagger c}|n\rangle_\alpha = e^{-2\alpha^*\alpha} \sum_{s=0}^{\min[j,n]} \frac{(-1)^s \sqrt{j!\,n!}\,(2\alpha)^{j-s}(2\alpha^*)^{n-s}}{(j-s)!\,(n-s)!\,s!}$$

$$D_{nk} = {}_\alpha\langle n|e^{i\pi c^\dagger c}|k\rangle_\alpha = e^{-2\alpha^*\alpha} \sum_{s=0}^{\min[n,k]} \frac{(-1)^s \sqrt{n!\,k!}\,(2\alpha)^{n-s}(2\alpha^*)^{k-s}}{(n-s)!\,(k-s)!\,s!}$$
(3.12)

If $\sum_{j,k=0}^{\infty}\sum_{n=0}^{\infty} D_{jn}D_{nk}|j\rangle_{\alpha\alpha}\langle k| = \mathbf{I}$, then its determinant $\det(\sum_{j,k=0}^{\infty}\sum_{n=0}^{\infty} D_{jn}D_{nk}|j\rangle_{\alpha\alpha}\langle k|) = 1$.

However in fact

$$\det\left(\sum_{j,k=0}^{M\to\infty}\sum_{n=0}^{M\to\infty}D_{jn}D_{nk}|j\rangle_{\alpha\alpha}\langle k|\right)=e^{-4\alpha^*\alpha M}\neq 1 \tag{3.13}$$

that Eq. (3.11$L$) $\neq$ Eq. (3.11$R$), the assumption is invalid! The closure relation of Fock states $\sum_n|n\rangle\langle n|=\mathbf{I}$ is refuted under $\mathbf{D}$ transformation.

The refutation of the closure relation of projection operators has decisive significance for the resolution of the EPR paradox! Because under the orthogonal normalization complete set, the premise that two non-commuting mechanical quantities cannot be simultaneously diagonalized is based on the closure relation of projection operators. $\sum_n|\varphi_n\rangle\langle\varphi_n|$ The assumption of $\mathbf{I}$. In other words, there is no sufficient reason to believe that two non-commuting mechanical quantities cannot be simultaneously diag onalized under the orthogonal normalization complete set! In fact, it can be strictly proven: $\sum_n|\varphi_n\rangle\langle\varphi_n|\neq\mathbf{I}$ At this time, mechanical quantities that do not commute can be strictly irreversibly diagonalized under the orthogonal normalized set.

## IV. DIAGONALIZATION OF ANGULAR MOMENTUM, COORDINATES, AND MOMENTUM

4.1 Angular momentum $J_x$, $J_y$ and $J_z$ are simultaneously diagonalized

According to Schwinger's angular momentum theory $J_x=(b^\dagger c+bc^\dagger)/2$, $J_y=(b^\dagger c-bc^\dagger)/2i$, $J_z=(b^\dagger b-c^\dagger c)/2$, taking Eq.(3.1) determines the orthogonal normalized basis $\mathbf{U}^H\mathbf{U}=\mathbf{I}$, then the angular momentum $J_\mu(\mu=x,y,z)$ and its square $J_\mu^2$ are strictly diagonalized as

$$\begin{aligned}\langle\varphi_m|J_\mu|\varphi_n\rangle&=\delta_{mn}j_\mu^{(1)}\\ \langle\varphi_m|J_\mu^2|\varphi_n\rangle&=\delta_{mn}j_\mu^{(2)}\end{aligned} \tag{4.1}$$

where $j_x^{(1)}=\frac{n}{2}\frac{\tau^*+\tau}{1+\tau^*\tau}$, $j_y^{(1)}=\frac{n}{2i}\frac{\tau^*-\tau}{1+\tau^*\tau}$, $j_z^{(1)}=\frac{n}{2}\frac{\tau^*\tau-1}{1+\tau^*\tau}$; $j_\mu^{(2)}=\frac{n}{4}+\frac{n-1}{n}\left(j_\mu^{(1)}\right)^2$.

It is easy to verify that at this time $J_x$, $J_y$ and $J_z$'s simultaneous diagonalization still satisfies the Robertson uncertainty relation Eq.(1.1) because

$$\begin{aligned}\langle\varphi_m|J_xJ_y|\varphi_n\rangle&=\delta_{mn}\left(\frac{n-1}{n}j_x^{(1)}j_y^{(1)}+\frac{i}{2}j_z^{(1)}\right)\\ \langle\varphi_m|J_yJ_x|\varphi_n\rangle&=\delta_{mn}\left(\frac{n-1}{n}j_x^{(1)}j_y^{(1)}-\frac{i}{2}j_z^{(1)}\right)\end{aligned} \tag{4.2}$$

$$\begin{aligned}\langle\varphi_m|J_yJ_z|\varphi_n\rangle&=\delta_{mn}\left(\frac{n-1}{n}j_y^{(1)}j_z^{(1)}+\frac{i}{2}j_x^{(1)}\right)\\ \langle\varphi_m|J_zJ_y|\varphi_n\rangle&=\delta_{mn}\left(\frac{n-1}{n}j_y^{(1)}j_z^{(1)}-\frac{i}{2}j_x^{(1)}\right)\end{aligned} \tag{4.3}$$

$$\begin{aligned}\langle\varphi_m|J_zJ_x|\varphi_n\rangle&=\delta_{mn}\left(\frac{n-1}{n}j_z^{(1)}j_x^{(1)}+\frac{i}{2}j_y^{(1)}\right)\\ \langle\varphi_m|J_xJ_z|\varphi_n\rangle&=\delta_{mn}\left(\frac{n-1}{n}j_z^{(1)}j_x^{(1)}-\frac{i}{2}j_y^{(1)}\right)\end{aligned} \tag{4.4}$$

Noting that $\Delta J_\mu=\sqrt{\langle\varphi_n|J_\mu^2|\varphi_n\rangle-(\langle\varphi_n|J_\mu|\varphi_n\rangle)^2}$, thus we have

$$\Delta J_x \Delta J_y = \sqrt{\frac{1}{n^2}\left(j_x^{(1)} j_y^{(1)}\right)^2 + \frac{1}{4}\left(j_z^{(1)}\right)^2} \geq \frac{1}{2}\left|\overline{[J_x, J_y]}\right| = \frac{1}{2}\left|j_z^{(1)}\right| \tag{4.5}$$

$$\Delta J_y \Delta J_z = \sqrt{\frac{1}{n^2}\left(j_y^{(1)} j_z^{(1)}\right)^2 + \frac{1}{4}\left(j_x^{(1)}\right)^2} \geq \frac{1}{2}\left|\overline{[J_y, J_z]}\right| = \frac{1}{2}\left|j_x^{(1)}\right| \tag{4.6}$$

$$\Delta J_z \Delta J_x = \sqrt{\frac{1}{n^2}\left(j_z^{(1)} j_x^{(1)}\right)^2 + \frac{1}{4}\left(j_y^{(1)}\right)^2} \geq \frac{1}{2}\left|\overline{[J_z, J_x]}\right| = \frac{1}{2}\left|j_y^{(1)}\right| \tag{4.7}$$

4.2 Simultaneous diagonalization of coordinates and momentum
4.2.1 Dark states and zero points of coordinates and momentum
Introducing the time reversal operator [Appendix 2]

$$\begin{aligned}\mathbf{T} &= \mathbf{T_1} e^{\frac{\pi}{2}(b^\dagger c - bc^\dagger)} \\ \mathbf{T^{-1}} &= e^{-\frac{\pi}{2}(b^\dagger c - bc^\dagger)} \mathbf{T_1^{-1}}\end{aligned} \tag{4.8}$$

Let the time-reversed state of $|\varphi_n\rangle$ be

$$|\varphi_n^T\rangle = \mathbf{T}|\varphi_n\rangle = \sum_{k=0}^{n} \frac{(C_n^k)^{1/2}(-\tau^*)^{n-k}}{(1+\tau^*\tau)^{n/2}}|n-k, k\rangle \tag{4.9}$$

Then **U**'s time reversal $\mathbf{W} = \mathbf{TU} = \text{diag}(|\varphi_0^T\rangle, \cdots, |\varphi_n^T\rangle)$ forms a new gauge $\mathbf{W^H W = I}$. Now let

$$\begin{aligned}a &\equiv \Theta c \Theta^{-1} = \frac{\tau^* b + c}{(1+\tau^*\tau)^{1/2}} \\ a^\dagger &\equiv \Theta c^\dagger \Theta^{-1} = \frac{\tau b^\dagger + c^\dagger}{(1+\tau^*\tau)^{1/2}}\end{aligned} \tag{4.10}$$

Satisfy $[a, a^\dagger] = 1$. Introduce the coordinate and momentum operators $\mathbf{q} = (a^\dagger + a)/\sqrt{2}$, $\mathbf{p} = i(a^\dagger - a)/\sqrt{2}$. It is easy to prove

$$\begin{aligned}\langle \varphi_m^T | a^\dagger a | \varphi_n^T \rangle &\equiv 0 \\ \langle \varphi_m | (a^\dagger a)^2 | \varphi_n \rangle &\equiv 0\end{aligned} \tag{4.11}$$

$$\begin{aligned}\langle \varphi_m^T | \mathbf{q} | \varphi_n^T \rangle &\equiv 0 \\ \langle \varphi_m | \mathbf{q}^2 | \varphi_n \rangle &\equiv \frac{1}{2}\delta_{mn}\end{aligned} \tag{4.12}$$

$$\begin{aligned}\langle \varphi_m^T | \mathbf{p} | \varphi_n^T \rangle &\equiv 0 \\ \langle \varphi_m | \mathbf{p}^2 | \varphi_n \rangle &\equiv \frac{1}{2}\delta_{mn}\end{aligned} \tag{4.13}$$

Eq.(4.11) The equation indicates : in the orthogonal normalized basis $\{|\varphi_n^T\rangle\}$, the eigenvalues of the number operator and its square are always zero. Without any observable particle number and fluctuations, we refer to this complete set of all $\{|\varphi_n^T\rangle, n = 0 \to \infty\}$ as the dark state of the particle number operator $a^\dagger a$. Eq.(4.12) and Eq.(4.13) indicate that the dark state is not only the infinitely degenerate eigenstate of the zero point of coordinates and momentum, but also the infinitely degenerate eigenstate of the sum of the squares of coordinates and momentum. The zero point vibrational energy of the vacuum state is merely one subset of the dark state zero point vibrational energy.

It is noted that the angular momentum operator $J_\mu$ is also strictly diagonalized in the dark state.

$$\langle\varphi_m^T|J_\mu|\varphi_n^T\rangle = -\delta_{mn} j_\mu^{(1)} \tag{4.14}$$

The dark state of the particle number operator $a^\dagger a$ is the non-degenerate eigenstate of angular momentum. At this time, fluctuations of coordinates and momentum

$$\begin{aligned}\Delta q &= \sqrt{\langle\varphi_n^T|\mathbf{q}^2|\varphi_n^T\rangle - (\langle\varphi_n^T|\mathbf{q}|\varphi_n^T\rangle)^2} = \frac{1}{\sqrt{2}} \\ \Delta p &= \sqrt{\langle\varphi_n^T|\mathbf{p}^2|\varphi_n^T\rangle - (\langle\varphi_n^T|\mathbf{p}|\varphi_n^T\rangle)^2} = \frac{1}{\sqrt{2}}\end{aligned} \tag{4.15}$$

The Robertson uncertainty relation for coordinates and momentum takes its minimum value in the dark state.

$$\Delta q \Delta p = \frac{1}{2} \tag{4.16}$$

4.2.2 The translation of dark states and the non-zero eigenvalues of coordinates and momentum

Introducing the translation operator

$$\mathfrak{D} = e^{\alpha a^\dagger - \alpha^* a} = \exp\left(\frac{\alpha \tau b^\dagger - \alpha^* \tau^* b}{\sqrt{1+\tau^*\tau}}\right)\exp\left(\frac{\alpha c^\dagger - \alpha^* c}{\sqrt{1+\tau^*\tau}}\right) \tag{4.17}$$

Denote the translation state of the dark state as

$$|\varphi_n^{DT}\rangle = \mathfrak{D}|\varphi_n^T\rangle = \frac{(a_T^\dagger)^n}{\sqrt{n!}} e^{\alpha a^\dagger - \alpha^* a}|0,0\rangle \tag{4.18}$$

where $a_T^\dagger = \mathbf{T}a^\dagger \mathbf{T}^{-1} = \frac{b^\dagger - \tau^* c^\dagger}{\sqrt{1+\tau^*\tau}}$.

Let $\mathbf{V} = \text{diag}(|\varphi_0^{DT}\rangle, \cdots, |\varphi_n^{DT}\rangle)$, then $\mathbf{V}^H\mathbf{V}=\mathbf{I}$ is a gauge under the translation of dark states. In this gauge, it is always true that

$$\begin{aligned}\langle\varphi_m^{DT}|a^k|\varphi_n^{DT}\rangle &= \delta_{mn}\alpha^k \\ \langle\varphi_m^{DT}|(a^\dagger)^k|\varphi_n^{DT}\rangle &= \delta_{mn}\alpha^{*k}\end{aligned} \tag{4.19}$$

$$\begin{aligned}\langle\varphi_m^{DT}|a^\dagger a|\varphi_n^{DT}\rangle &= \delta_{mn}\alpha^*\alpha \\ \langle\varphi_m^{DT}|(a^\dagger a)^2|\varphi_n^{DT}\rangle &= \delta_{mn}\alpha^*\alpha(\alpha^*\alpha + 1)\end{aligned} \tag{4.20}$$

$$\begin{aligned}\langle\varphi_m^{DT}|\mathbf{q}^k|\varphi_n^{DT}\rangle &= \delta_{mn}f_k(q) \\ \langle\varphi_m^{DT}|\mathbf{p}^k|\varphi_n^{DT}\rangle &= \delta_{mn}f_k(p)\end{aligned} \tag{4.21}$$

Here: $f_k(q) = \frac{e^{-q^2}}{2^k}\frac{d^k}{dq^k}e^{q^2}$, $f_k(p) = \frac{e^{-p^2}}{2^k}\frac{d^k}{dp^k}e^{p^2}$, $q = \frac{\alpha^*+\alpha}{\sqrt{2}}$, $p = i\frac{\alpha^*-\alpha}{\sqrt{2}}$

At this time, the fluctuations of coordinates and momentum

$$\begin{aligned}\Delta q &= \sqrt{\langle\varphi_n^{DT}|\mathbf{q}^2|\varphi_n^{DT}\rangle - (\langle\varphi_n^{DT}|\mathbf{q}|\varphi_n^{DT}\rangle)^2} = \frac{1}{\sqrt{2}} \\ \Delta p &= \sqrt{\langle\varphi_n^{DT}|\mathbf{p}^2|\varphi_n^{DT}\rangle - (\langle\varphi_n^{DT}|\mathbf{p}|\varphi_n^{DT}\rangle)^2} = \frac{1}{\sqrt{2}}\end{aligned} \tag{4.22}$$

In the translational state of the dark state, the Robertson uncertainty relation still takes the minimum value: $\Delta q \Delta p = \frac{1}{2}$

## V. SUMMARY

(1) Reversible diagonalization gives the first type motion constant; irreversible diagonalization

gives the second type motion constant. The second type motion constant under the irreversible diagonalization of the hidden order parameter is associated with the quantum phase transition of the system. Models such as Dick model, TC model, and SB model can be strictly solved analytically under irreversible diagonalization and provide the corresponding scaling behavior.

(2) The EPR paradox ultimately stems from the closure relation of the projection operator $\mathbf{UU^H} = \sum_n |\varphi_n\rangle\langle\varphi_n| = \mathbf{I}$. The assumption of $\mathbf{I}$. It is only true under the orthogonal normalized basis in Abelian groups ($\mathbf{UU^H = U^H U}$)! At this time, two mechanical quantities that do not commute generally cannot be simulta neously reversibly diagonalized, leading to paradoxes such as coordinates and momentum not being able to simultaneously constitute physical reality.

(3) The falsifiability of the closure relation of projection operators under orthogonal normalized basis reveals that the wave function expressed by Dirac's vector has the non-commutativity of a non-Abelian group ($\mathbf{UU^H \neq U^H U}$). In the orthogonal normalized set containing hidden parameters, two mechanical quantities that do not commute can be simultaneously irreversibly diagonalized. In short: The EPR paradox is merely a dilemma of physical reality under reversible diagonalization, which is resolved under irreversible diagonalization of physical reality.

**Appendix 1**: Proof that 'Two non-commuting mechanical quantities cannot have a common eigenstate in an orthonormal complete set.'
It is known that the mechanical quantities A and B satisfy the commutation relation:
$$[A, B] \neq 0.$$
Assuming they have a common eigenstate in the orthogonal normalized set $\langle\varphi_m|\varphi_n\rangle = \delta_{mn}$ under which: $A|\varphi_n\rangle = \alpha_n|\varphi_n\rangle$, $B|\varphi_n\rangle = \beta_n|\varphi_n\rangle$. Take any state $|\psi\rangle \neq 0$, and perform
$$[A, B]|\psi\rangle = (AB - BA)|\psi\rangle.$$
According to the expansion principle of quantum mechanical states, we have
$$|\psi\rangle = \sum_n |\varphi_n\rangle\langle\varphi_n||\psi\rangle = \sum_n c_n|\varphi_n\rangle.$$
Thus, we have
$$[A, B]|\psi\rangle = (AB - BA)|\psi\rangle = \sum_n c_n(AB - BA)|\varphi_n\rangle.$$
From the assumption, it must follow that
$$AB|\varphi_n\rangle = A\beta_n|\varphi_n\rangle = \alpha_n\beta_n|\varphi_n\rangle, \quad BA|\varphi_n\rangle = B\alpha_n|\varphi_n\rangle = \alpha_n\beta_n|\varphi_n\rangle.$$

Substituting into the previous expression gives
$$[A, B]|\psi\rangle = (AB - BA)|\psi\rangle = \sum_n c_n(\alpha_n\beta_n - \alpha_n\beta_n)|\varphi_n\rangle \equiv 0.$$
Since $|\psi\rangle$ is arbitrary and non-zero, it must be $[A, B] = 0$. However, this contradicts the premise $[A, B] \neq 0$, therefore the assumption is invalid. Two mechanical quantities that do not commute cannot have a common eigenstate in an orthonormal complete set.

A key step in the above proof $|\psi\rangle = \sum_n |\varphi_n\rangle\langle\varphi_n||\psi\rangle$ There is a fundamental premise, namely the projection operator of quantum mechanics symbol $\sum_n |\varphi_n\rangle\langle\varphi_n|$ is an identity operator ($\sum_n |\varphi_n\rangle\langle\varphi_n| \equiv I$) for any state . Otherwise, the proposition cannot be empirically verified!

**Appendix 2: Introduction of the time reversal operator**

$$\mathbf{T} \equiv \mathbf{T}_1 e^{\frac{\pi}{2}(b^\dagger c - bc^\dagger)}.$$

where $\mathbf{T}_1$ is regarded as the classical time reversal operator, which only acts on the time parameter t and the imaginary unit i

$$\mathbf{T}_1 t \mathbf{T}_1^{-1} = -t, \quad \mathbf{T}_1 i \mathbf{T}_1^{-1} = -i,$$

and $e^{\frac{\pi}{2}(b^\dagger c - bc^\dagger)}$ is regarded asquantum time reversal operator, it only acts on the creation and annihilation operators $b^\dagger, b, c^\dagger, c$ acts

$$e^{\frac{\pi}{2}(b^\dagger c - bc^\dagger)} b^\dagger e^{-\frac{\pi}{2}(b^\dagger c - bc^\dagger)} = -c^\dagger, \quad e^{\frac{\pi}{2}(b^\dagger c - bc^\dagger)} b e^{-\frac{\pi}{2}(b^\dagger c - bc^\dagger)} = -c,$$

$$e^{\frac{\pi}{2}(b^\dagger c - bc^\dagger)} c^\dagger e^{-\frac{\pi}{2}(b^\dagger c - bc^\dagger)} = b^\dagger, \quad e^{\frac{\pi}{2}(b^\dagger c - bc^\dagger)} c e^{-\frac{\pi}{2}(b^\dagger c - bc^\dagger)} = b.$$

for the bosonic operator $a = \frac{\tau^* b + c}{(1+\tau^*\tau)^{\frac{1}{2}}}$ and $a^\dagger = \frac{\tau b^\dagger + c^\dagger}{(1+\tau^*\tau)^{\frac{1}{2}}}$, whose time reversal is

$$a_T = \mathbf{T} a \mathbf{T}^{-1} = \frac{b - \tau c}{(1+\tau^*\tau)^{1/2}},$$

$$a_T^\dagger = \mathbf{T} a^\dagger \mathbf{T}^{-1} = \frac{b^\dagger - \tau^* c}{(1+\tau^*\tau)^{1/2}}.$$

the time reversal of the corresponding coordinates and momentum is

$$\mathbf{q}_T = \mathbf{T} \mathbf{q} \mathbf{T}^{-1} = \frac{a_T^\dagger + a_T}{\sqrt{2}},$$

$$\mathbf{p}_T = \mathbf{T} \mathbf{p} \mathbf{T}^{-1} = -i\frac{a_T^\dagger - a_T}{\sqrt{2}}.$$

The time reversal of coordinates remains invariant; the time reversal of momentum changes sign under the condition of invariance. The time reversal of angular momentum is

$$\mathbf{T} J_\mu \mathbf{T}^{-1} = -J_\mu \quad (\mu = x, y, z).$$